\documentclass[12pt,aps,prd,preprint,tightenlines,superscriptaddress,
   showpacs,nofootinbib]{revtex4-1}
\usepackage{graphicx}

\usepackage{amsmath}

\newcommand{\PRE}[1]{{#1}} 

\newcommand{\del}{\partial}
\newcommand{\nbox}{{\,\lower0.9pt\vbox{\hrule \hbox{\vrule height 0.2 cm
\hskip 0.2 cm \vrule height 0.2 cm}\hrule}\,}}

\newcommand{\ipb}{\text{pb}^{-1}}
\newcommand{\ifb}{\text{fb}^{-1}}

\newcommand{\mev}{\text{MeV}}
\newcommand{\gev}{\text{GeV}}
\newcommand{\tev}{\text{TeV}}

\newcommand{\be}{\begin{equation}}
\newcommand{\ee}{\end{equation}}
\newcommand{\bea}{\begin{eqnarray}}
\newcommand{\eea}{\end{eqnarray}}
\newcommand{\lsim}{\lower.7ex\hbox{$\;\stackrel{\textstyle<}{\sim}\;$}}
\newcommand{\gsim}{\lower.7ex\hbox{$\;\stackrel{\textstyle>}{\sim}\;$}}

\begin{document}

\preprint{UH-511-1170-2011}
\preprint{UCI-TR-2011-11}

\title{
\PRE{\vspace*{1.3in}}
Collider Searches for Fermiophobic Gauge Bosons
\PRE{\vspace*{0.3in}}
}

\author{Joseph Bramante}
\affiliation{Department of Physics and Astronomy, University of
Hawai'i, Honolulu, HI 96822, USA
\PRE{\vspace*{.1in}}
}

\author{R. S. Hundi}
\affiliation{Department of Physics and Astronomy, University of
Hawai'i, Honolulu, HI 96822, USA
\PRE{\vspace*{.1in}}
}
\affiliation{Department of Theoretical Physics,
Indian Association for the Cultivation of Science,
2A $\&$ 2B Raja S.C. Mullick Road,
Kolkata - 700 032, India
\PRE{\vspace*{.1in}}
}

\author{Jason Kumar}
\affiliation{Department of Physics and Astronomy, University of
Hawai'i, Honolulu, HI 96822, USA
\PRE{\vspace*{.1in}}
}

\author{Arvind Rajaraman%
\PRE{\vspace*{.4in}}
}
\affiliation{Department of Physics and Astronomy, University of California, Irvine, CA  92697
\PRE{\vspace*{.5in}}
}

\author{David Yaylali}
\affiliation{Department of Physics and Astronomy, University of
Hawai'i, Honolulu, HI 96822, USA
\PRE{\vspace*{.1in}}
}


\begin{abstract}
\PRE{\vspace*{.3in}} We explore the phenomenology of an extra U(1) gauge boson
which primarily couples to standard model gauge bosons.
We classify all possible parity-odd couplings up to dimension 6 operators.  We
then study the prospects for the
detection of such a boson at the LHC and show that the electroweak decay
channels lead to very clean signals, allowing us to probe couplings
well into the TeV scale.
\end{abstract}

\pacs{
12.60.-i 
14.70.Pw  
}
\maketitle

\section{Introduction}

One of the most natural possibilities for
physics beyond the standard model is the existence of new gauge groups.
In particular, new U(1) gauge groups which are Higgsed at the TeV scale
can lead to new massive gauge bosons (see e.g.\cite{Leike:1998wr,Rizzo:2006nw,Langacker:2008yv,Nath:2010zj}). Such massive
gauge bosons are a generic feature of many extensions
of the standard model like grand unified theories~\cite{Ross:1985ai}. String
theoretic constructions can also lead to a plethora of new gauge
groups~\cite{Blumenhagen:2000vk,Cvetic:2001nr,Cvetic:2002wh,Uranga:2003pz,Cvetic:2004nk,Marchesano:2004yq,Cvetic:2004xx,Kumar:2005hf,Kumar:2007zza,Douglas:2006xy}.

The new gauge bosons
can couple to
the standard model in many ways. Usually they are assumed
to have direct couplings to the standard
model fermions, and they can then be directly produced as resonances in
colliders.
There has been great interest in collider searches for such $Z'$
gauge bosons, and strong constraints have been placed on such
resonances~\cite{Alcaraz:2006mx,Jaffre:2009dg,Chatrchyan:2011wq}.

A more interesting possibility is if the new gauge boson has
no direct couplings to the standard model fermions
(we will refer to such a gauge boson as being {\it fermiophobic}).
The new gauge boson (hereafter referred to as $X$)
may then have loop-induced couplings to the standard model if there are
fermions charged under both the new gauge group and the standard model.
If the fermions  are very heavy, then it may be kinematically
impossible to produce them on-shell; they would instead be integrated out
to yield effective higher dimensional operators coupling $X$ to
standard model gauge bosons. We will focus here on
this possibility.

There are several scenarios for
a fermiophobic $X$. One commonly studied possibility
is that of kinetic
mixing~\cite{Holdom:1985ag,delAguila:1995rb,Dienes:1996zr,Kumar:2006gm,Feldman:2006wb,Chang:2006fp,Chang:2007ki,Feldman:2007wj},
in which there is a
dimension 4 operator which mixes the kinetic terms
of $X$ and the hypercharge gauge boson.  This kinetic mixing induces
suppressed couplings between $X$ and the standard model fermions, and
the $X$ then appears as a $Z'$ with a small coupling.
There are, however, many models where such a kinetic
mixing term is absent; for example if  the heavy fermions are
coupled to a non-Abelian standard model group, then the kinetic mixing diagrams
are forbidden. Effective operators must then couple $X$ to at least two standard
model gauge bosons~\cite{Anastasopoulos:2006cz}.  We would then need to search for $X$ through its couplings
to two gauge bosons.

If the $X$ couples only to electroweak gauge bosons,
$X$ can be produced at hadron colliders
through vector boson fusion,
followed by the decay $X \rightarrow ZZ \rightarrow 4l$.
This possibility was considered in~\cite{Kumar:2007zza},
where the authors considered
a fermiophobic gauge boson coupled
to electroweak gauge bosons through dimension
6 operators.   This was further
extended in~\cite{Antoniadis:2009ze}, where it was pointed out that $X$ can
couple to electroweak gauge bosons  through dimension 4 operators as well, enhancing
the production cross section.

Here we consider the more general case
where $X$ couples both to gluons as well as to
electroweak gauge bosons (as would happen if  the heavy fermions couple to SU(3)$_{QCD}$ as well as SU(2)$_L$).
We examine the prospects for an LHC search for a massive spin-1 boson
coupled to gluons and electroweak gauge bosons through
all possible parity-odd couplings up to dimension 6.
We find that the on-shell  production of
$X$ arises through a unique dimension 6 operator
coupling $X$ to gluons.
However, there is greater freedom in
writing operators coupling $X$ to electroweak gauge bosons.
As a result, decay can arise through a variety of dimension 4 and 6 operators, the coefficients
of which determine the branching fraction to the final states $ZZ$, $Z\gamma$ and $W^+ W^-$.
Interestingly, $X$ cannot decay to $\gamma \gamma$.  This follows from the Landau-Yang
theorem \cite{LandauYang},
which asserts that a massive spin-1 boson cannot decay to two massless vector bosons.

The organization of the paper is as
follows. In section II we present the effective operator description of the coupling of the
hidden sector gauge boson to standard model gauge bosons.  In section III we describe
our analysis of LHC detection prospects for this signal, assuming $\sqrt{s}=7~\tev$.
We conclude with a discussion of our results in section IV.

\section{Effective theory of the fermiophobic gauge boson}
We consider a theory with a new gauge group U(1)$_X$ spontaneously broken by the expectation value of a charged scalar field $\Phi$,
which is eaten by the Higgs mechanism giving the gauge boson
$X$ a mass.
We will consider
the case where the gauge boson $X$ has only negligible couplings to
standard model fermions, but couples nontrivially to standard model gauge bosons.
We will further specialize to the case where $X$ is a pseudovector; the vector
case will be considered elsewhere.

SU(3) gauge invariance constrains the coupling of $X$ to gluons to be a
combination of
three effective operators:
\bea
\begin{aligned}
{\cal O}_{Xgg}^1  &= {1\over \Lambda^2} \epsilon^{\mu\rho \alpha \beta }X_{\mu}D^\nu
G_{\alpha \nu}^a  G_{\beta \rho}^a
\\
{\cal O}_{Xgg}^2  &= {1\over \Lambda^2} \epsilon^{\mu\rho \alpha \beta }\del^\nu X_{\mu}
G_{\alpha \nu}^a  G_{\beta \rho}^a
\\
{\cal O}_{Xgg}^3  &= {1\over \Lambda^2} \epsilon^{\alpha \beta \nu \rho}\del_\mu X^{\mu}
G_{\alpha \beta}^a G_{\nu \rho}^a,
\end{aligned}
\eea
where $D_\mu$ is a covariant derivative and $G_{\alpha \beta}^a$ is a gluon field strength.
The operator ${\cal O}_{Xgg}^3 $ cannot
contribute to any process where the $X$ is on-shell, since the momentum of $X$ is
orthogonal to its physical polarizations.  Thus we can ignore this term if the narrow-width
approximation is valid (and we will find that it is).
${\cal O}_{Xgg}^2 $ also cannot contribute to any process where the $X$ is on-shell.
One can see this by assuming without loss of generality that $X$ is in the rest frame
($p_X=(M_X,0,0,0)$) with  polarization $\epsilon_X =(0,1,0,0)$. The only nonvanishing
terms are thus
$\epsilon^{1\rho \alpha \beta } \del^0 X_{1}  G_{\alpha 0}^a  G_{\beta \rho}^a$,
and it is easy to verify that this expression will vanish due to the antisymmetric property
of the epsilon tensor.

The only operator which contributes to on-shell production
of $X$ is
${\cal O}_{Xgg}^1$.
The corresponding vertex for this operator is
\bea
\Gamma_{\mu \nu \rho} ^{Xgg} (k_X ,k_1 ,k_2 )&=&
{1\over \Lambda^2} \left[\epsilon_{\mu \nu \rho \sigma}(-k_1^2 k_2^\sigma +k_2^2 k_1^\sigma )
+ \epsilon_{\mu  \rho \sigma \tau} k_{1\nu} k_2^\sigma k_1^\tau
- \epsilon_{\mu  \nu \sigma \tau} k_{2\rho} k_2^\sigma k_1^\tau \right].
\eea
Note that in this case the vertex is only nonvanishing if at least one gluon is off-shell.
This is a consequence of the Landau-Yang Theorem.

Since electroweak symmetry is broken, it is not necessary for operators
to exactly satisfy the SU(2)$_L$ Ward Identity.  As a result, we may write operators
in the effective
Lagrangian in terms of the $Z$ and $W$ gauge fields as well as the field strengths.
The most general $XZZ$ coupling can be derived from 4 effective operators (see also
\cite{Keung:2008ve}):
\bea
\begin{aligned}
{\cal O}_{XZZ}^1 &= \epsilon^{\mu\nu\rho\sigma}X_{\mu}Z_\nu Z_{\rho\sigma} =
\epsilon^{\mu\nu\rho\sigma} {X_{\mu} H^\dagger D_\nu H Z_{\rho\sigma} \over |H|^2} \\
{\cal O}_{XZZ}^2 &= {1\over \Lambda^2} \epsilon^{\mu\rho \alpha \beta }X_{\mu}
\del^\nu Z_{\alpha \nu}  Z_{\beta \rho} \\
{\cal O}_{XZZ}^3 &= {1\over \Lambda^2} \epsilon^{\mu \rho \alpha \beta } \del^\nu
X_{\mu}Z_{\alpha \nu}  Z_{\beta \rho} \\
{\cal O}_{XZZ}^4 &= {1\over \Lambda^2} \epsilon^{\alpha \beta \rho \sigma}
\del_\mu X^{\mu}Z_{\alpha \beta} Z_{\rho \sigma},
\end{aligned}
\eea
where $Z_{\alpha \beta}$ is the $Z$-boson field strength.

Using the same arguments as for the gluon coupling, it is clear that
${\cal O}_{XZZ}^3$ and ${\cal O}_{XZZ}^4$ cannot contribute to any process involving
an on-shell $X$.
The vertices for the other two effective operators are
\bea
\begin{aligned}
\Gamma_{\mu \nu \rho}^{XZZ,1} (k_X ,k_1 ,k_2 ) &=
\epsilon_{\mu \nu \rho \sigma}( k_2^\sigma - k_1^\sigma ) \\
\Gamma_{\mu \nu \rho}^{XZZ,2} (k_X ,k_1 ,k_2 ) &=
{1\over \Lambda^2} \left[\epsilon_{\mu \nu \rho \sigma}(-k_1^2 k_2^\sigma +k_2^2 k_1^\sigma )
+ \epsilon_{\mu  \rho \sigma \tau} k_{1\nu} k_2^\sigma k_1^\tau
- \epsilon_{\mu  \nu \sigma \tau} k_{2\rho} k_2^\sigma k_1^\tau \right]
\end{aligned}
\eea
Note that if the $Z$s are on-shell, as we require, the dimension 4 operator yields
the same vertex as the dimension 6 operator:
\bea
\Gamma_{\mu \nu \rho}^{XZZ,2} \approx  -{M_Z^2 \over \Lambda^2} \Gamma_{\mu \nu \rho}^{XZZ,1}.
\eea
Thus we need only consider the dimension 6 operator in the remainder of this paper.  In the case where
interactions are mediated by a dimension 4 operator, the coupling of
$X$ to electroweak states can be easily obtained using the expression above.

The $XZ\gamma$ vertex does not have a symmetry between the two field strengths. For the photon
only the field strength can appear, while the field $Z_{\mu}$ can appear by itself. The
most general such interaction is a combination of the operators
\bea
\begin{aligned}
{\cal O}_{XZ\gamma }^1 &= \epsilon^{\mu\nu\rho\sigma}X_{\mu}Z_\nu F_{\rho\sigma}
\\
{\cal O}_{XZ\gamma }^2 &= {1\over \Lambda^2}\epsilon^{\mu\rho \alpha \beta} \del^\nu X_{\mu}
(Z_{\alpha \nu}  F_{\beta \rho} + F_{\alpha\nu}  Z_{\beta \rho})
\\
{\cal O}_{XZ\gamma }^3 &= {1\over \Lambda^2}\epsilon^{\mu\rho \alpha \beta} \del^\nu X_{\mu}
(Z_{\alpha \nu}  F_{\beta \rho} - F_{\alpha\nu}  Z_{\beta \rho})
\\
{\cal O}_{XZ\gamma }^4 &= {1\over \Lambda^2}\epsilon^{\mu\rho \alpha \beta}X_{\mu}
\del^\nu Z_{\alpha \nu}  F_{\beta \rho}
\\
{\cal O}_{XZ\gamma }^5 &= {1\over \Lambda^2}\epsilon^{\mu\rho \alpha \beta }X_{\mu}
\del^\nu F_{\alpha \nu}  Z_{\beta \rho}
\\
{\cal O}_{XZ\gamma }^6 &= {1\over \Lambda^2}\epsilon^{\alpha \beta \nu \rho}X^{\mu}
\del_\mu Z_{\alpha \beta} F_{\nu \rho}
\\
{\cal O}_{XZ\gamma }^7 &= {1\over \Lambda^2}\epsilon^{\alpha \beta \nu \rho}
\del_\mu X^{\mu} Z_{\alpha \beta}  F_{\nu \rho}
\end{aligned}
\eea
where $F_{\alpha \beta}$ is an electromagnetic field strength.  The
operators ${\cal O}_{XZ\gamma }^2$,
${\cal O}_{XZ\gamma }^5$ and ${\cal O}_{XZ\gamma }^7$ do not contribute to any
process in which $X$ and the photons are on-shell.

We can further assume that the only operators we generate are at most dimension 6
when written in manifestly
SU(2)-covariant notation.  In this case, the only electroweak operators we can write are
\bea
\begin{aligned}
{\cal O}^1 &= {C_1\over \Lambda^2} \epsilon^{\mu\rho \alpha \beta }X_{\mu} \mbox{Tr}[
\del^\nu C_{\alpha \nu}  C_{\beta \rho}]
\\
{\cal O}^2 &= {C_2\over 2 \Lambda^2} \epsilon^{\mu\rho \alpha \beta }X_{\mu}\del^\nu
B_{\alpha \nu}  B_{\beta \rho}
\end{aligned}
\eea
where $C$ is the SU(2) gauge field strength, and $B$ is the hypercharge field strength.

These operators then completely determine the vertices for $XZZ$, $XZ\gamma$ $XWW$ and
$X\gamma \gamma$
(for on-shell $X$).  Defining
\bea
\Gamma_{\mu \nu \rho}(k_X ,k_1, k_2 ) &=&
(k_{2 \rho} \epsilon_{\mu \nu \sigma \tau} k_1^{\sigma} k_2^{\tau} -
k_{1 \nu}\epsilon_{\mu \rho \sigma \tau} k_1^{\sigma} k_2^{\tau}
+ \epsilon_{\mu \nu \rho \sigma} k_1^{\sigma} k_2 \cdot k_2
- \epsilon_{\mu \nu \rho \sigma} k_2^{\sigma} k_1 \cdot k_1   ), \nonumber
\eea
we have
\bea
\Gamma_{\mu \nu \rho}^{XZZ} (k_X ,k_1 ,k_2  )  &=&  {1 \over \Lambda^2}(C_1 \cos^2 \theta_W
+ C_2 \sin^2 \theta_W) \Gamma_{\mu \nu \rho}(k_X ,k_1,k_2 ) \\
\Gamma_{\mu \nu \rho}^{XZ \gamma } (k_X ,k_1 ,k_2  )  &=&   {1 \over \Lambda^2}
(C_1 - C_2) \sin \theta_W \cos \theta_W
\Gamma_{\mu \nu \rho}(k_X ,k_1,k_2) \\
\Gamma_{\mu \nu \rho}^{XW^{+} W^{-} } (k_X ,k_1 ,k_2  )  &=&  {C_1 \over \Lambda^2}
\Gamma_{\mu \nu \rho}(k_X ,k_1,k_2) \\
\Gamma_{\mu \nu \rho}^{X\gamma \gamma } (k_X ,k_1 ,k_2  )  &=&  {1 \over \Lambda^2}
(C_1 \sin^2 \theta_W + C_2 \cos^2 \theta_W)
\Gamma_{\mu \nu \rho}(k_X ,k_1,k_2).
\eea
If all particles are on-shell, these vertices simplify considerably;
\bea
\Gamma_{\mu \nu \rho}^{XZZ} (k_X ,k_1 ,k_2  ) &=& {M_Z^2 \over \Lambda^2}
(C_1 \cos^2 \theta_W + C_2 \sin^2 \theta_W)
\epsilon_{\mu \nu \rho \sigma}( k_1^\sigma - k_2^\sigma )
\\
\Gamma_{\mu \nu \rho} ^{XZ\gamma } (k_X ,k_1 ,k_2  ) &=&{M_Z^2 \over \Lambda^2}
(C_2 - C_1) \sin \theta_W \cos \theta_W
\epsilon_{\mu \nu \rho \sigma}k_2^\sigma
\\
\Gamma_{\mu \nu \rho}^{XWW} (k_X ,k_1 ,k_2 ) &=&
C_1 {M_W^2 \over \Lambda^2}\epsilon_{\mu \nu \rho \sigma}( k_1^\sigma - k_2^\sigma ) \\
\Gamma_{\mu \nu \rho}^{X\gamma \gamma } (k_X ,k_1 ,k_2 ) &=& 0.
\eea

\section{$X$ Production and Decay}

We will be considering processes in which the $X$ boson is produced on-shell
in hadron collisions. As we have seen, the Landau-Yang theorem prohibits the decay of a
massive spin-1 particle to two
massless vector particles and also prohibits resonance production of a massive spin-1
particle from two massless vectors.
QCD processes therefore always
produce the $X$ boson in association with  a jet.
Note that this is only true for on-shell production of $X$; if $X$ is not on-shell,
it can be produced without extra jets.
For the moment we neglect this possibility;  it would be interesting to see if
off-shell production of $X$ can lead to nontrivial results.

The parton-level process $gg \rightarrow Xg$ also vanishes.
The only relevant parton-level production channels are therefore
$qg \rightarrow qX$, $\bar q g \rightarrow \bar q X$ and $q\bar q \rightarrow gX$; see Fig. 1.

\begin{figure}[!h]
\centering
  \includegraphics[width=2in]{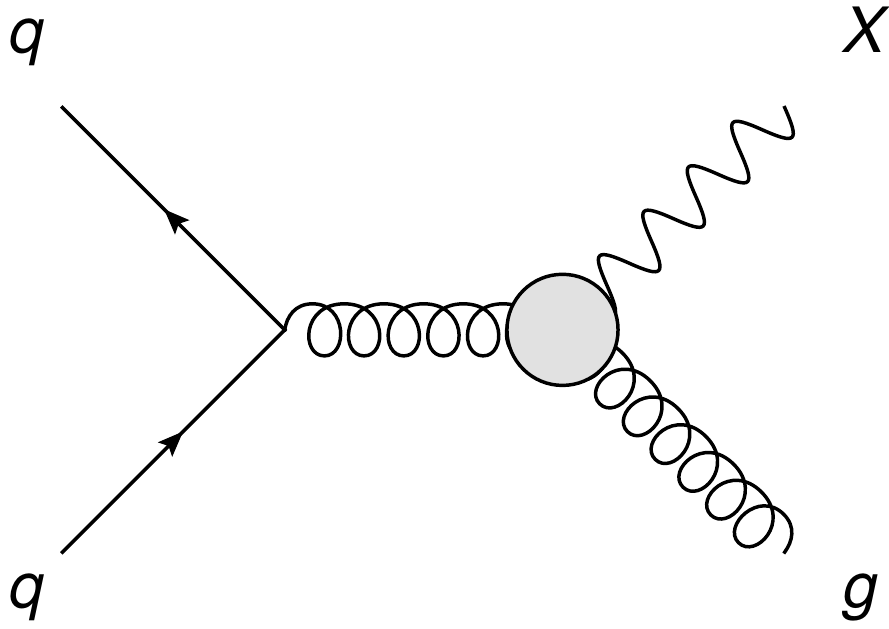}
  \caption{$X$ production through $ qg $, $ \bar{q} g $, and $ q \bar{q} $.}
\end{figure}

The branching fractions for $X$ decay can also be calculated. The branching fraction
for $X \rightarrow gg$ and $X \rightarrow ggg$ turn out to be zero.
As a result, the only hadronic decay of $X$ to fewer than four jets is through the process
$X \rightarrow g q \bar q$.  Depending on the relative values of the coefficients for the
gluon and electroweak operators, this can be an important decay channel.

In this paper we are interested in the electroweak decay channels only.  For the purposes of
illustrating relative branching fractions to these channels, we will assume that the operator coefficients
are chosen such that the partial width for
$X \rightarrow g q \bar q$ is negligible. (Our final result will be independent of this assumption.)  In
this case the primary decay modes are $ZZ$, $Z\gamma$ and $WW$.  We find
\bea
\begin{aligned}
\Gamma(X\rightarrow W W)
&=  (42~\mev) \left({\tev \over \Lambda}\right)^4
\left({M_X\over \tev}\right)^3 \left(1-{4M_W ^2 \over M_X ^2}\right)^{5/2}
 C_1^2
\\
\Gamma(X\rightarrow ZZ) &=
(16~\mev) \left({\tev \over \Lambda}\right)^4
\left({M_X\over \tev}\right)^3 \left(1-{4M_Z ^2 \over M_X ^2}\right)^{5/2}
(C_1 + C_2 \tan^2 \theta_W)^2 \label{eq:BR} \\
\Gamma(X\rightarrow \gamma Z) &=
(4.9~\mev) \left({\tev \over \Lambda}\right)^4
\left({M_X\over \tev}\right)^3
\left(1-{M_Z^2 \over M_X ^2}\right)^{3} \left(1+{M_Z ^2 \over M_X ^2} \right)
(C_2 - C_1 )^2.
\end{aligned}
\eea
Note that for $M_X, \Lambda \sim \tev$, the decay width of $X$ is indeed much
smaller than its mass, justifying our use of the narrow-width approximation.

\begin{figure}[!h]
\includegraphics*[width=3in]{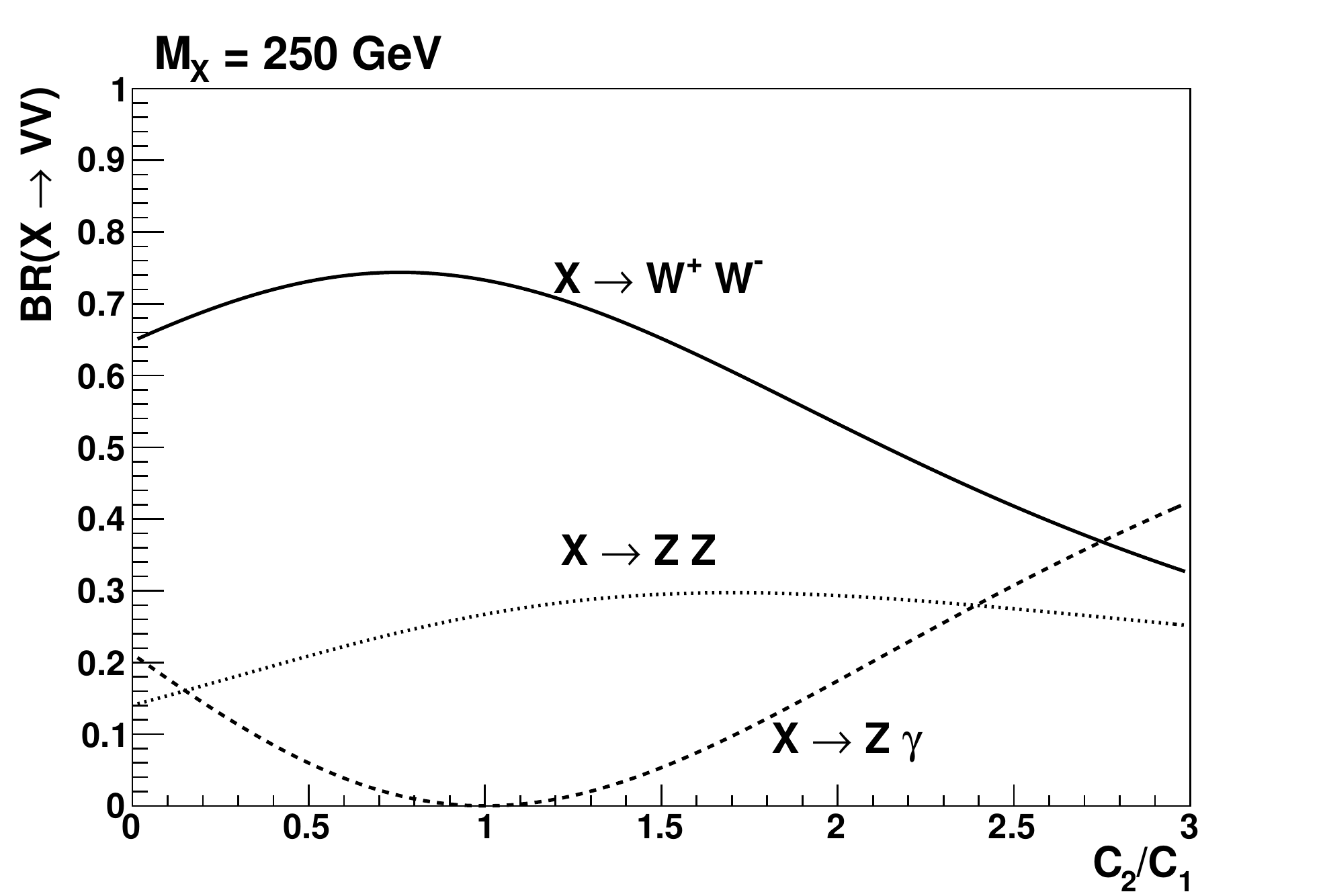}
\includegraphics*[width=3in]{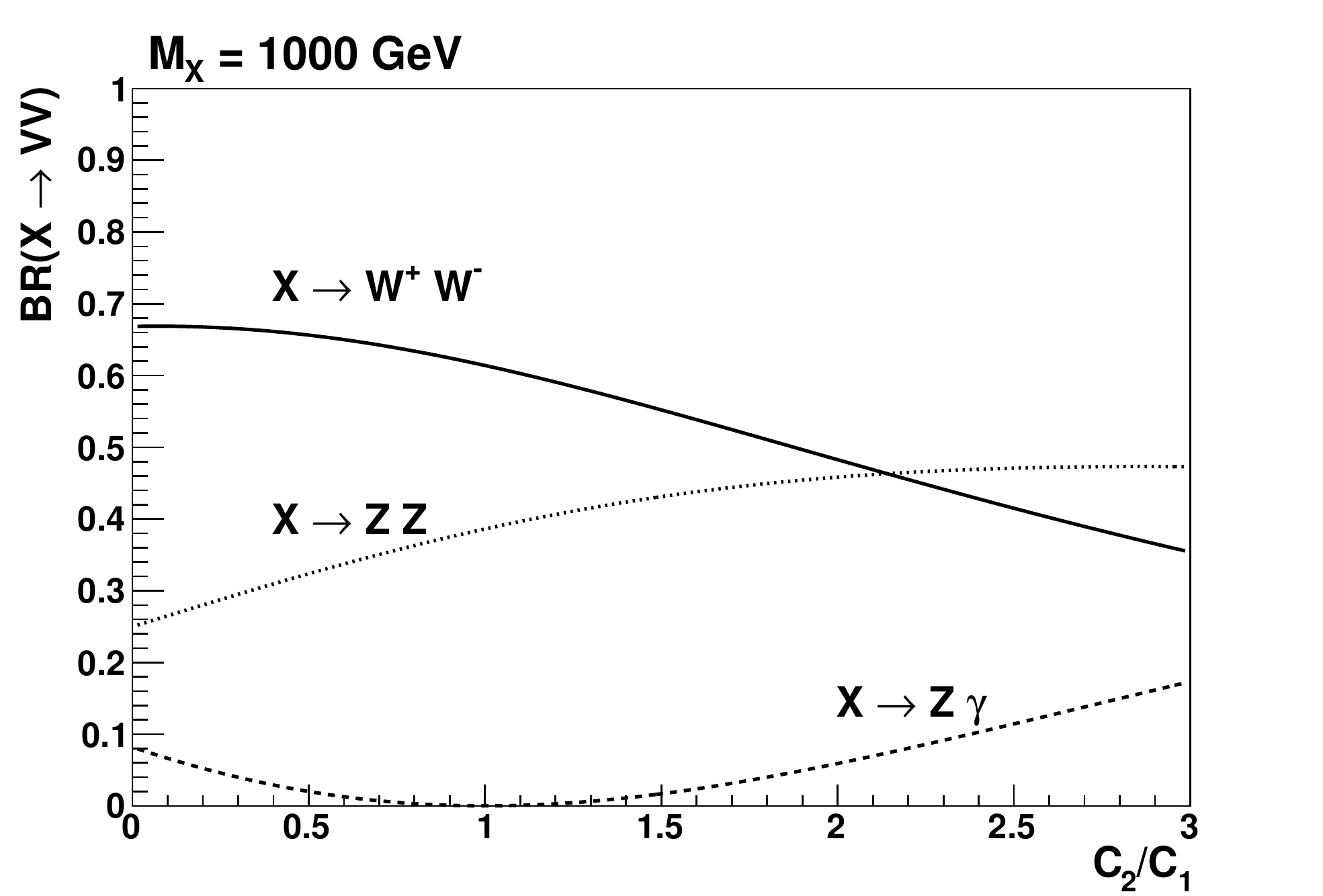}
\caption{Branching ratios for $X$ decaying to standard model electroweak gauge bosons for $M_X=250~\gev$ (left panel)
and for $M_X=1000~\gev$ (right panel).  We have assumed that the branching fraction to $g q \bar q$ is negligible.}
\label{fig:BR}
\end{figure}

In Fig.~\ref{fig:BR} we plot the branching fractions BR$(X \rightarrow ZZ, W^+ W^- , Z\gamma)$
as a function of $C_2 / C_1$ for
$M_X = 250~\gev$ and $M_X = 1000~\gev$.

\section{Collider Analysis}

In this analysis we will
study potential signals at the 7 TeV LHC.
We will focus on the case of $X$ production through QCD couplings via the
operator
\bea
{\cal O}_{Xgg} &=& {\cal O}_{Xgg}^1  = {1\over \Lambda^2} \epsilon^{\mu\rho \alpha \beta }X_{\mu}D^\nu
G_{\alpha \nu}^a  G_{\beta \rho}^a
\eea
followed by $X \rightarrow ZZ$ and $X \rightarrow Z\gamma$ decays, which are the cleanest.
We will further specialize to the  case where the $Z$ decays to leptons.
We have simulated the signal and standard model background in \textsc{madgraph 5}~\cite{Alwall:2011uj}, showered the partons
using \textsc{pythia 6.4.22}~\cite{Sjostrand:2006za}, and
performed a detector simulation in \textsc{pgs4}~\cite{PGS4}.
We consider each final state separately.

\subsection{Cuts}

(a) \textit{ZZ decays}:
For $X\to ZZ$ decays the  signal is 4 leptons plus a jet. The primary background is $ZZ$ + jet
production.  We impose the following cuts:
\begin{itemize}
\item{One jet with $p_{T} \geq 50~\gev $, $| \eta | < 2.5$}
\item{4 leptons with $p_{T} \geq 20~\gev $, $| \eta | < 2.5$, and pairwise invariant masses
in the range 80-100~\gev}
\end{itemize}

(b) \textit{Z$\gamma$ decays}:
For $X \to Z\gamma$ decays the signal is 2 leptons, a photon and a jet.  The primary background
is $Z\gamma$ + jet production.  We impose the following cuts:
\begin{itemize}
\item{One jet with $p_{T} \geq 50~\gev $, $| \eta | < 2.5$}
\item{2 leptons with $p_{T} \geq 20~\gev $, $| \eta | < 2.5$, and invariant mass
in the range 80-100~\gev}
\item{1 photon with $p_{T} \geq 10~\gev $}
\end{itemize}

\begin{figure}[!h]
  \centering
  \includegraphics[width=3.2in,height=2.5in]{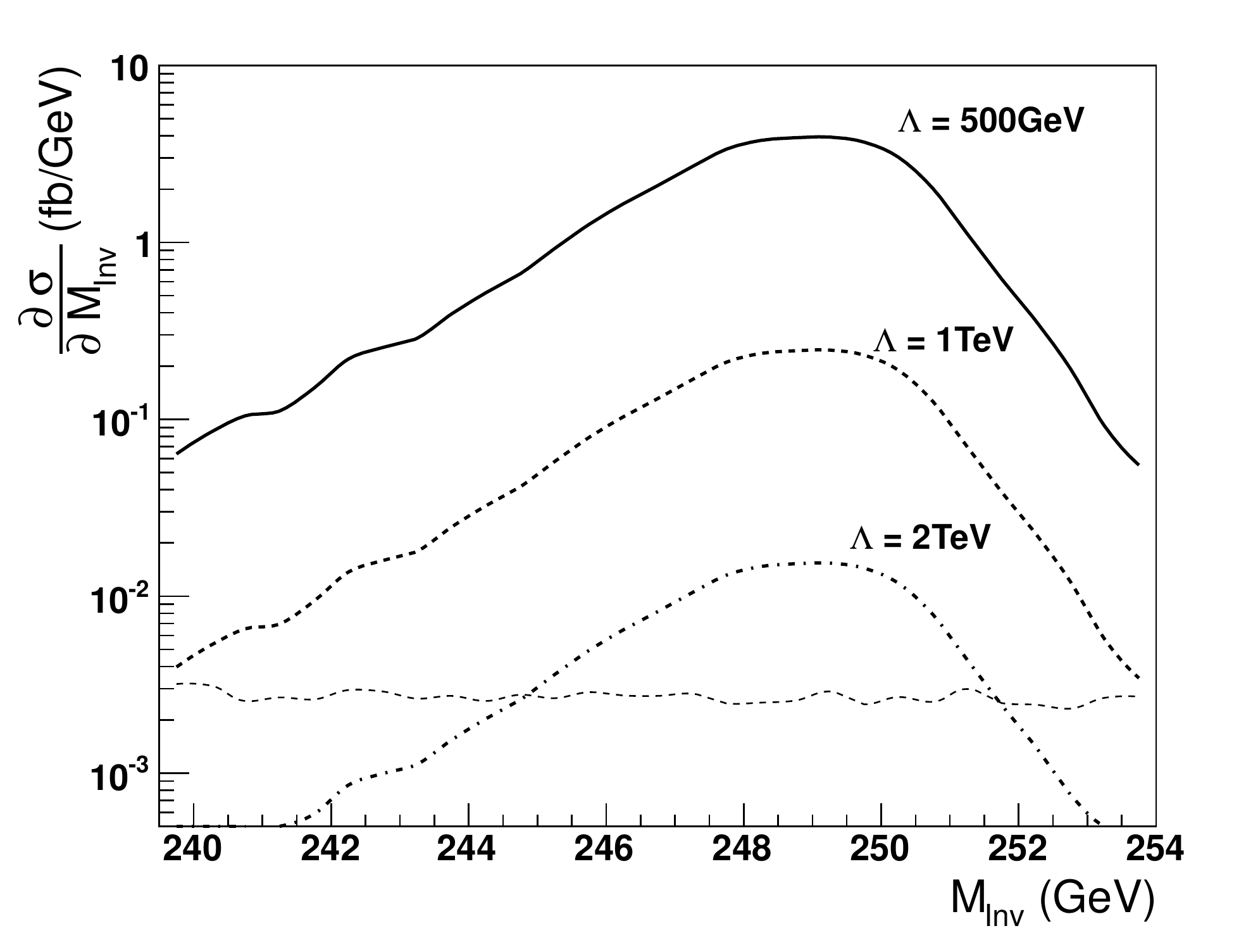}
  \includegraphics[width=3.2in,height=2.5in]{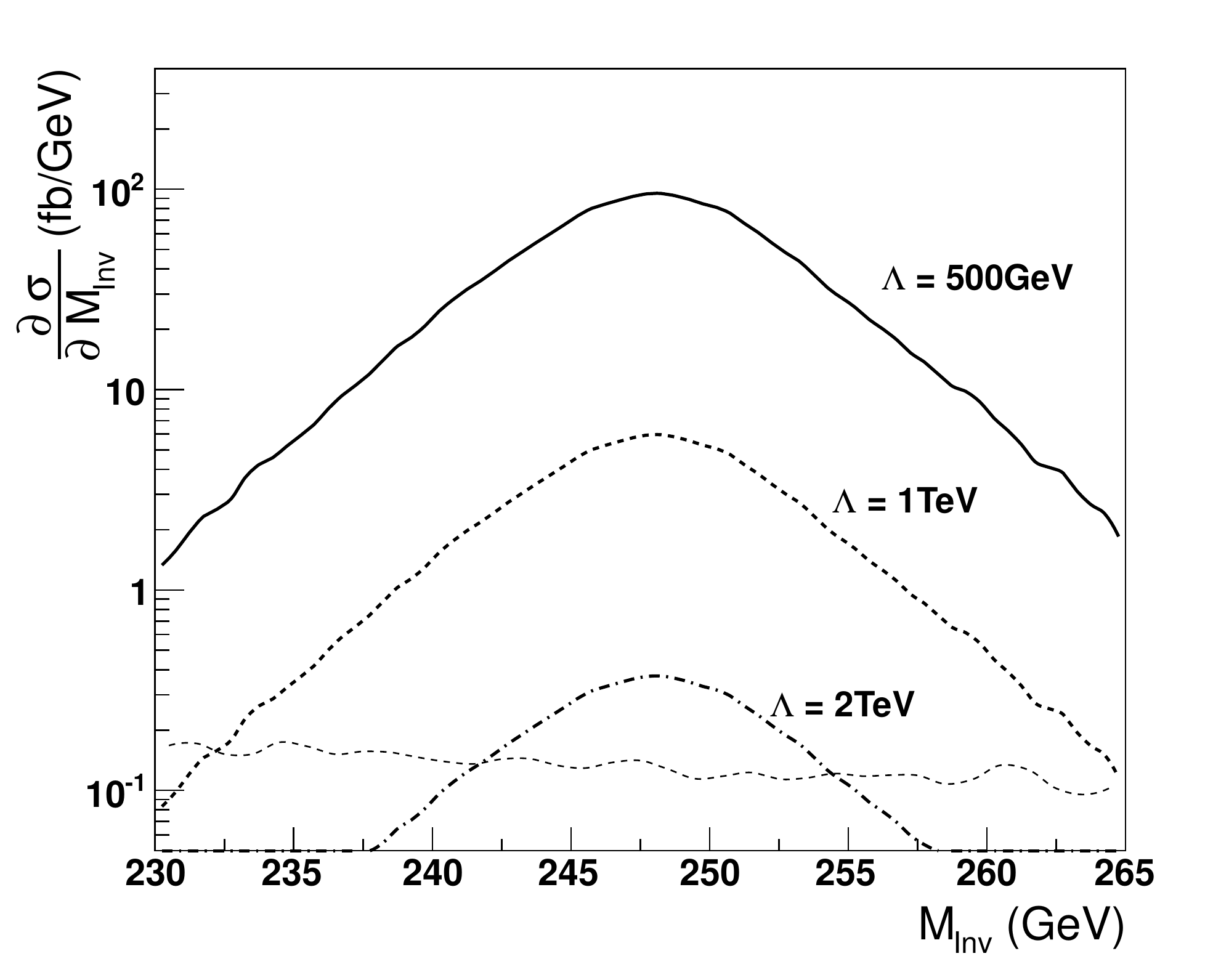}
  \caption{ \label{fig:SigBG}Invariant mass spectrum for signal ($M_X = 250~\gev$) and background for
  the LHC at $\sqrt{s}=7~\tev$,
  assuming the only electroweak coupling is through operator ${\cal O}^1$.
  Left panel: Signal and background for $X \rightarrow ZZ$, for different values
  of $\Lambda$.
  Right panel: Signal and background for $X \rightarrow Z\gamma$.  Both
  signal and background cross sections
  are generally lower for the $ZZ$ process due to the extra factor of
  the dilepton branching ratio.}
\end{figure}

To look for the $X$ resonance, we can study the total invariant mass of
the 4 leptons (or 2 leptons and photon).
The invariant mass distributions for the signal vs. background (assuming the only electroweak
coupling is through operator ${\cal O}^1$) are shown in
Fig.~\ref{fig:SigBG} for $M_{X} = 250~\gev$ and for various values of $\Lambda$.
The standard model background events give a smooth distribution over the relevant
invariant mass combinations (see also~\cite{Baur:1988cq,Baur:1989cm}).
The cross sections for the signal are well above background for $\Lambda$ as high as 2 TeV.

Since signal events will exhibit a narrow invariant mass peak, our analysis will compare the
number of observed events to the number of expected background events with an invariant
mass within $\pm 10\% $ of a given central value $m_{central}$.
For both $ZZ$ and $Z\gamma$ channels, this invariant mass cut drastically lowers
background cross sections.
In Table~\ref{table:br} we present the signal and standard model background cross sections
for events satisfying the cuts with
4 lepton (or 2 lepton plus photon) invariant
mass within 10\% of the given $m_{central}$.

\begin{table}[htbp]
\caption{Table of signal and standard model background production cross sections (in fb) for the LHC at
$\sqrt{s} =7~\tev$, for 4 leptons and for 2 leptons and a photon, as labeled.
The signal cross sections are normalized by taking $\Lambda / B(X\rightarrow VV)^{1\over 4} =~\tev$
for each final state.  We assume the cuts
described in the text and the assumption that the invariant mass is within 10\% of $m_{\text{central}}$.  }
\begin{tabular}{@{\extracolsep{15pt}}  l  c  c  c  c }
\hline
\hline

& $\sigma_{\text{BG}} $(fb) & $\sigma_{\text{BG}} $(fb) & $\sigma_{\text{sig}}/ \text{BR}(ZZ) $ (fb) & $\sigma_{\text{sig}}/ \text{BR}(Z \gamma) $ (fb) \\
$m_{\text{central}}$ (GeV) & $ pp\rightarrow j l^{+}l^{-}l^{+}l^{-}$ & $pp\rightarrow j \gamma l^{+}l^{-}$
& $ pp\rightarrow j l^{+}l^{-}l^{+}l^{-}$ & $pp\rightarrow j \gamma l^{+}l^{-}$\\
\hline
250 & 0.26 & 6.4 & 17.8 & 690 \\
500 & 0.050 & 0.76 & 5.47 & 141 \\
750 & 0.010 & 0.17 & 1.27 & 36\\
1000 & 0.0021 & 0.034 & 0.26 & 9.6\\
1250 & 0.0004 & 0.014 & 0.054 & 2.53\\
1500 &0.0001 & 0.0051 & 0.012 & 0.66\\
1750 & $<$0.0001 & $<$0.0010 & 0.0032 & 0.18\\
2000 & $<$0.0001 & $<$0.0010 & 0.0008 & 0.049\\
\hline
\hline
\end{tabular}
\label{table:br}
\end{table}

\subsection{Detection Prospects}

We find the number $B$ of
background events with $4 l$ ($2l + \gamma$) invariant mass within
$\pm 10\%$ of any given $m_{central}$ and compare this to the
number $S$ of signal events within the same invariant mass window,
assuming $M_X = m_{central}$.
The significance is defined as $s={S\over \sqrt{B}}$.
For each point in parameter space, we can find the
luminosity required to achieve discovery.  When the number of expected background events at
a certain luminosity is less than one, we define discovery as $S\geq 5$; otherwise, we define
discovery as $s \geq 5$.  For all of the parameter space considered one finds $S / B \geq 0.2$
at discovery.

In a realistic experimental analysis the actual signal significance would be reduced by
a trials factor associated with the freedom in choosing $m_{central}$, the center of the
invariant mass analysis window.

\begin{figure}[!h]
\centering
  \includegraphics[width=3.2in]{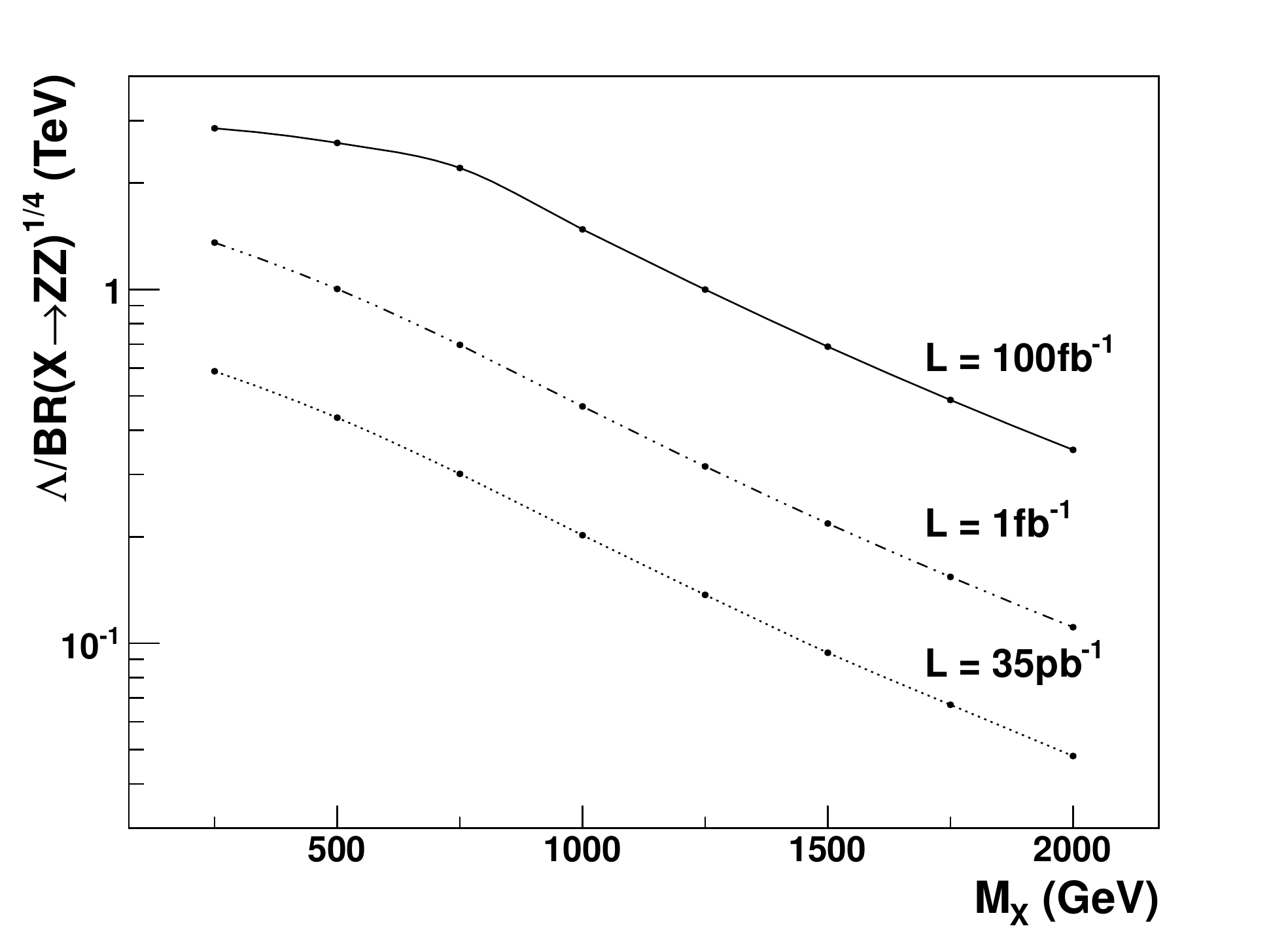}
  \includegraphics[width=3.2in]{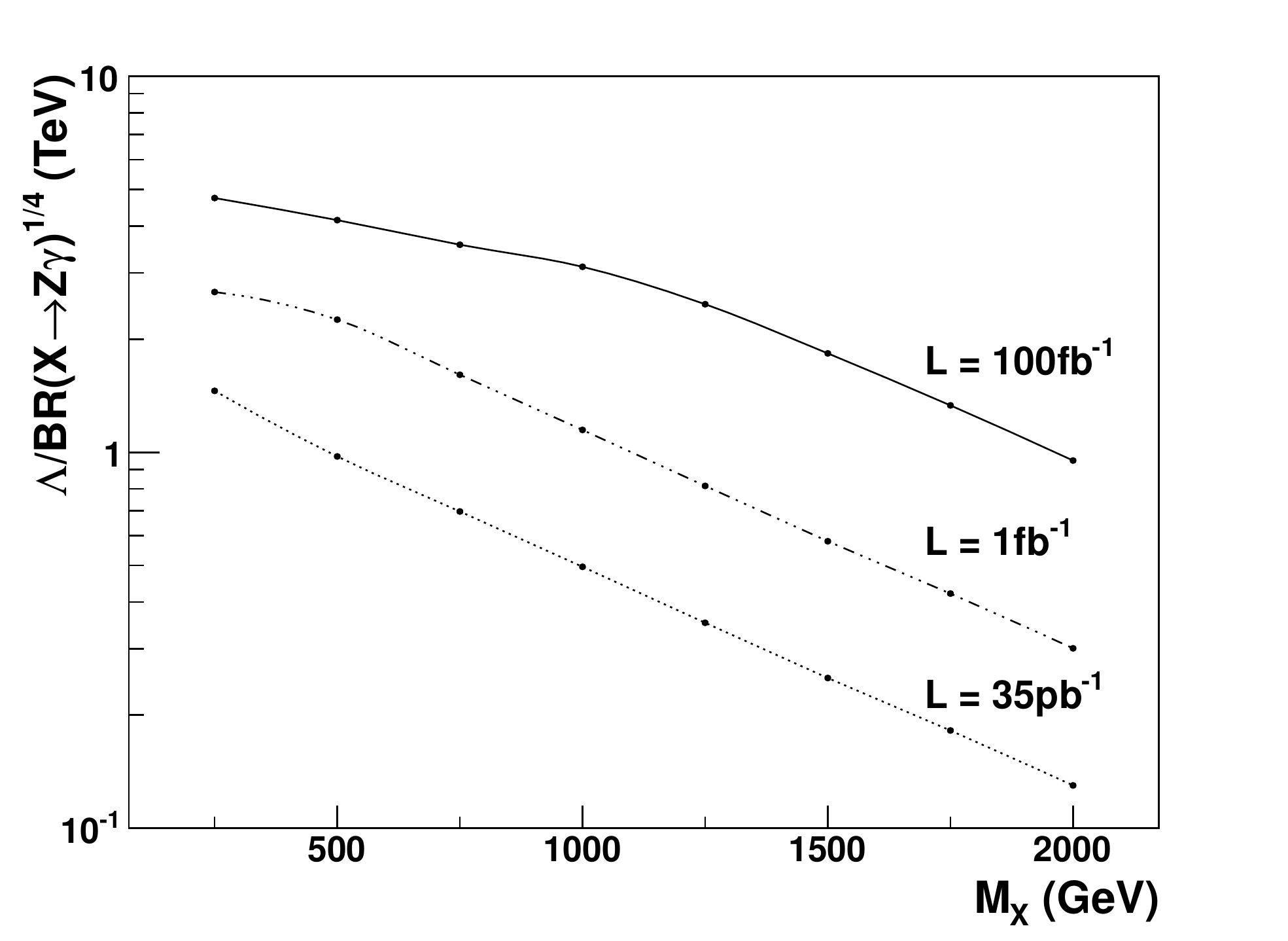}
  \caption{\label{fig:LHCreach}Discovery reach of the LHC at $\sqrt{s}=7~\tev$ for the
  $X \rightarrow ZZ$ channel (left panel) and the $X \rightarrow Z \gamma$ channel (right panel).
  $\Lambda$ is the mass scale of the dimension 6 operator coupling $X$ to gluons. }
\end{figure}

Since we have seen that the narrow-width approximation is valid for $X$, the detection prospects
of the LHC depend on the electroweak coupling operator coefficients ($C_1$ and $C_2$) only through the
branching fraction for $X$ to decay to each channel.  Since the minimum cross section for discovery scales as
the production cross section times the branching ratio,
\begin{align}
\sigma_{pp \rightarrow X+jet \rightarrow VVjet} &= \sigma_{prod} \times \text{BR}(X \rightarrow VV) \\
&\propto \Lambda^{-4} \times \text{BR}(X \rightarrow VV),
\end{align}
we define the mass reach in terms of the quantity ${\Lambda}/[\text{BR}(X \rightarrow VV)]^{1\over 4}$.  This mass reach is then
independent of the relative strengths of couplings to the strong and electroweak sectors, as encoded
by the factors $C_1$ and $C_2$. We plot the mass reach accessible at the LHC for various luminosities at collider energy
7 TeV in both the $ZZ$ and $Z\gamma$ channel in Fig.~\ref{fig:LHCreach}.

Note that, for large $M_X$ and ${\cal L} = 35~\ipb$, the discovery reach can drop as
low as $\Lambda \sim 40~\gev$.  For typical models, this would imply that particles
which have been integrated out to generate the higher-dimensional effective operator
are in fact lighter than the energy of the hard process, rendering the effective
operator analysis inconsistent.  Moreover, one might expect additional operators with
dimension greater than 6 to provided contributions which are suppressed by additional powers of
$M_Z^2 / \Lambda^2$; if this factor is large, then one cannot ignore the effect of these
contributions.  While these issues would hold with typical models, one can also have
models in which the degrees of freedom which have been integrated out are indeed heavy,
but have a large multiplicity which serves to reduce $\Lambda$ (see also~\cite{Kumar:2007zza}).
Moreover, the contribution
of operators with dimension $> 6$ depends on the details of the heavy degrees of freedom.
Thus, we have plotted the discovery reach even if $\Lambda < M_{Z,X}$ for completeness,
but in those cases the effective operator analysis may not be valid.

\section{Conclusions}

We see that the best detection prospects arise from the $X \rightarrow Z\gamma$ channel, because
the $X\rightarrow ZZ$ channel suffers from the small branching fraction for two $Z$s
to decay to
4 leptons.  In both channels standard model backgrounds become
significant
for relatively
light $M_X$ and large luminosities.  As expected, the sensitivity of the LHC to
resonant $X$ production is greatly
enhanced when production through QCD processes is possible.  Comparing to the
results in~\cite{Kumar:2007zza},
we see that if $M_X = 1000~\gev$, then gluon couplings allow the LHC to probe operators suppressed
by a mass scale $\Lambda$ which is 10 times larger than the scale which could be probed if
only electroweak couplings were allowed.  Note again that our analysis has focused on a 7 TeV center
of mass energy; detection prospects would be improved
significantly if the center of mass energy were upgraded to $\sqrt{s} = 14~\tev$.  For example,
for $M_X = 1000~\gev$, a $100~\ifb$ LHC run at $\sqrt{s}=14~\tev$ will have roughly twice the mass reach
of a run at 7 TeV (in either the $ZZ$ or $Z\gamma$ channels).

It is worth noting that a resonance which decays to $ZZ$ and $W^+ W^-$ is a  characteristic
signature of
a relatively heavy Higgs boson.  A new pseudovector coupling to standard model gauge bosons can
thus ``counterfeit" standard Higgs signals~\cite{Fox:2011qc}.  It may be especially difficult to
distinguish these
possibilities, since a heavy Higgs decays to light fermions with a very small
branching fraction.  The
features of a pseudovector which can be used to distinguish it from a Higgs include the absence of
the  $\gamma \gamma$ channel and the fact that production through QCD processes requires
the presence of
an additional jet.  It would be interesting to study in detail the prospects for distinguishing
the spin of any resonance which couples to standard model gauge bosons.

It is also worth noting that we have focused only on effective coupling operators which are nontrivial
when the $X$ and the standard model gauge bosons are both on-shell.  If these operators vanish (or have
very small couplings), then the production and decay of $X$ may be dominated by operators which only
yield nontrivial vertices when $X$ is off-shell.  This would imply that $X$ production
is not associated
with a resonance peak.  A detailed collider study of such operators would be very interesting.
\\

\textbf{Acknowledgments}

We are grateful to C.~Duhr, B.~Thomas, J.~Wells and especially J.~Alwall for useful discussions.
The work of J.~K., R.~S.~H. and D.~Y. is supported
in part by Department of Energy grant DE-FG02-04ER41291.
The work of A.~R. is supported
in part by  NSF Grants PHY-0653656 and PHY-0709742.


\begin{thebibliography}{99}

\bibitem{Leike:1998wr}
  A.~Leike,
  Phys.\ Rept.\  {\bf 317}, 143 (1999)
  [arXiv:hep-ph/9805494].

\bibitem{Rizzo:2006nw}
  T.~G.~Rizzo,
  arXiv:hep-ph/0610104.

\bibitem{Langacker:2008yv}
  P.~Langacker,
  Rev.\ Mod.\ Phys.\  {\bf 81}, 1199 (2009)
  [arXiv:0801.1345 [hep-ph]].

\bibitem{Nath:2010zj}
  P.~Nath {\it et al.},
  Nucl.\ Phys.\ Proc.\ Suppl.\  {\bf 200-202}, 185 (2010)
  [arXiv:1001.2693 [hep-ph]].

\bibitem{Ross:1985ai}
  G.~G.~Ross,
{\it Frontiers In Physics,} 60 (Benjamin/Cummings, Reading, MA, 1984), p. 497.


\bibitem{Blumenhagen:2000vk}
  R.~Blumenhagen, L.~Goerlich, B.~Kors and D.~Lust,
  Fortsch.\ Phys.\  {\bf 49}, 591 (2001)
  [arXiv:hep-th/0010198].
\bibitem{Cvetic:2001nr}
  M.~Cvetic, G.~Shiu and A.~M.~Uranga,
   ``Chiral four-dimensional N=1 supersymmetric type 2A orientifolds from
  Nucl.\ Phys.\  B {\bf 615}, 3 (2001)
  [arXiv:hep-th/0107166].
\bibitem{Cvetic:2002wh}
  M.~Cvetic, P.~Langacker and G.~Shiu,
   ``A Three family standard - like orientifold model: Yukawa couplings and
  Nucl.\ Phys.\  B {\bf 642}, 139 (2002)
  [arXiv:hep-th/0206115].
\bibitem{Uranga:2003pz}
  A.~M.~Uranga,
   ``Chiral four-dimensional string compactifications with intersecting
  Class.\ Quant.\ Grav.\  {\bf 20}, S373 (2003)
  [arXiv:hep-th/0301032].
\bibitem{Cvetic:2004nk}
  M.~Cvetic, P.~Langacker, T.~j.~Li and T.~Liu,
  Nucl.\ Phys.\  B {\bf 709}, 241 (2005)
  [arXiv:hep-th/0407178].
\bibitem{Marchesano:2004yq}
  F.~Marchesano and G.~Shiu,
  Phys.\ Rev.\  D {\bf 71}, 011701 (2005)
  [arXiv:hep-th/0408059].
\bibitem{Cvetic:2004xx}
  M.~Cvetic and T.~Liu,
   ``Supersymmetric standard models, flux compactification and moduli
  Phys.\ Lett.\  B {\bf 610}, 122 (2005)
  [arXiv:hep-th/0409032].
\bibitem{Kumar:2005hf}
  J.~Kumar and J.~D.~Wells,
  JHEP {\bf 0509}, 067 (2005)
  [arXiv:hep-th/0506252].
\bibitem{Kumar:2007zza}
  J.~Kumar, A.~Rajaraman and J.~D.~Wells,
  Phys.\ Rev.\  D {\bf 77}, 066011 (2008)
  [arXiv:0707.3488 [hep-ph]].
\bibitem{Douglas:2006xy}
  M.~R.~Douglas and W.~Taylor,
  JHEP {\bf 0701}, 031 (2007)
  [arXiv:hep-th/0606109].

\bibitem{Alcaraz:2006mx}
  J.~Alcaraz {\it et al.} [ ALEPH and DELPHI and L3 and OPAL and LEP Electroweak Working Group Collaborations ],
  [hep-ex/0612034].

\bibitem{Jaffre:2009dg}
  M.~Jaffre  [CDF and D0 Collaboration],
   ``Search for high mass resonances in dilepton, dijet and diboson final states
  PoS E {\bf PS-HEP2009}, 244 (2009)
  [arXiv:0909.2979 [hep-ex]].

\bibitem{Chatrchyan:2011wq}
  S.~Chatrchyan {\it et al.}  [CMS Collaboration],
   ``Search for Resonances in the Dilepton Mass Distribution in $pp$ Collisions
  JHEP {\bf 1105}, 093 (2011)
  [arXiv:1103.0981 [hep-ex]].



\bibitem{Holdom:1985ag}
  B.~Holdom,
  Phys.\ Lett.\  B {\bf 166}, 196 (1986).


\bibitem{delAguila:1995rb}
  F.~del Aguila, M.~Masip and M.~Perez-Victoria,
  Nucl.\ Phys.\  B {\bf 456}, 531 (1995)
  [arXiv:hep-ph/9507455].

\bibitem{Dienes:1996zr}
  K.~R.~Dienes, C.~F.~Kolda and J.~March-Russell,
  Nucl.\ Phys.\  B {\bf 492}, 104 (1997)
  [arXiv:hep-ph/9610479].
\bibitem{Kumar:2006gm}
  J.~Kumar and J.~D.~Wells,
  Phys.\ Rev.\  D {\bf 74}, 115017 (2006)
  [arXiv:hep-ph/0606183].
\bibitem{Feldman:2006wb}
  D.~Feldman, Z.~Liu and P.~Nath,
   ``The Stueckelberg $Z$ Prime at the LHC: Discovery Potential, Signature
  JHEP {\bf 0611}, 007 (2006)
  [arXiv:hep-ph/0606294].
\bibitem{Chang:2006fp}
  W.~F.~Chang, J.~N.~Ng and J.~M.~S.~Wu,
  Phys.\ Rev.\  D {\bf 74}, 095005 (2006)
  [Erratum-ibid.\  D {\bf 79}, 039902 (2009)]
  [arXiv:hep-ph/0608068].
\bibitem{Chang:2007ki}
  W.~F.~S.~Chang, J.~N.~Ng and J.~M.~S.~Wu,
  Phys.\ Rev.\  D {\bf 75}, 115016 (2007)
  [arXiv:hep-ph/0701254].
\bibitem{Feldman:2007wj}
  D.~Feldman, Z.~Liu and P.~Nath,
   ``The Stueckelberg Z-prime Extension with Kinetic Mixing and Milli-Charged
  Phys.\ Rev.\  D {\bf 75}, 115001 (2007)
  [arXiv:hep-ph/0702123].

\bibitem{Anastasopoulos:2006cz}
  P.~Anastasopoulos, M.~Bianchi, E.~Dudas and E.~Kiritsis,
  JHEP \textbf{0611}, 057 (2006)
  [arXiv:hep-th/0605225].


\bibitem{Antoniadis:2009ze}
  I.~Antoniadis, A.~Boyarsky, S.~Espahbodi, O.~Ruchayskiy, J.~D.~Wells,
  Nucl.\ Phys.\  \textbf{B824}, 296-313 (2010).
  [arXiv:0901.0639 [hep-ph]].

  \bibitem{LandauYang}
  L.~D.~Landau, Dokl. Akad. Nawk., USSR {\bf 60}, 207 (1948)\\
  C.~N.~Yang, Phys. Rev. {\bf 77}, 242 (1950)


\bibitem{Keung:2008ve}
  W.~Y.~Keung, I.~Low and J.~Shu,
  Phys.\ Rev.\ Lett.\  {\bf 101}, 091802 (2008)
  [arXiv:0806.2864 [hep-ph]].

\bibitem{Alwall:2011uj}
  J.~Alwall, M.~Herquet, F.~Maltoni, O.~Mattelaer, T.~Stelzer,
  [arXiv:1106.0522 [hep-ph]].

\bibitem{Sjostrand:2006za}
  T.~Sjostrand, S.~Mrenna and P.~Z.~Skands,
  JHEP {\bf 0605}, 026 (2006)
  [arXiv:hep-ph/0603175].

\bibitem{PGS4}
PGS -- Pretty Good Simulator,
http://www.physics.ucdavis.edu/$\sim$conway/research/software/
pgs/pgs4-general.htm.

\bibitem{Baur:1989cm}
  U.~Baur and E.~W.~N.~Glover,
  Nucl.\ Phys.\  B \textbf{339}, 38 (1990).



\bibitem{Baur:1988cq}
  U.~Baur, E.~W.~N.~Glover and J.~J.~van der Bij,
  Nucl.\ Phys.\  B \textbf{318}, 106 (1989).


\bibitem{Fox:2011qc}
  P.~J.~Fox, D.~Tucker-Smith, N.~Weiner,
  [arXiv:1104.5450 [hep-ph]].




\end{thebibliography}
\end{document}